\documentclass[twocolumn,showpacs,amsmath,amssymb,10pt]{revtex4}
\usepackage{graphicx}
\usepackage{dcolumn}
\usepackage{stmaryrd}

\newcommand{\DelH}{\Delta\!_{\rm f}\!H^{\minuso}}
\newcommand{\DelG}{\Delta\!_{\rm f}\!G^{\minuso}}

\begin{document}

\title{Combined first--principles
  calculation and neural--network correction approach
  as a powerful tool in computational physics and chemistry
}

\author{LiHong Hu, XiuJun Wang, LaiHo Wong and GuanHua Chen}
 \email{ghc@everest.hku.hk}
 \homepage{http://yangtze.hku.hk}
\affiliation{Department of Chemistry, The University of Hong Kong,
Hong Kong, China}

\date{\today; submitted to Phys. Rev. Lett.}

\begin{abstract}
Despite of their success, the results of first-principles quantum
mechanical calculations contain inherent numerical errors caused by
various approximations. We propose here
a neural-network algorithm to greatly reduce these inherent errors.
As a demonstration, this combined quantum mechanical
calculation and neural-network correction
approach is applied to the evaluation of standard heat of formation
$\DelH$ and standard Gibbs energy of formation $\DelG$
for 180 organic molecules at 298 K.
A dramatic reduction of numerical errors is clearly shown with
systematic deviations being eliminated.
For examples, the root--mean--square
deviation of the calculated $\DelH$ ($\DelG$) for the
180 molecules is reduced
from 21.4 (22.3) kcal$\cdotp$mol$^{-1}$ to 3.1 (3.3)
kcal$\cdotp$mol$^{-1}$ for B3LYP/6-311+G({\it d,p})
and from 12.0 (12.9) kcal$\cdotp$mol$^{-1}$
to 3.3 (3.4) kcal$\cdotp$mol$^{-1}$ for B3LYP/6-311+G(3{\it df},2{\it p})
before and after the neural-network correction.

\end{abstract}

\pacs{31.15.Ew, 31.30.-i, 31,15,-p, 31.15.Ar} \maketitle

 One of the Holy Grails of computational science is to
quantitatively predict
properties of matters prior to experiments.
Despite the facts that the first-principles quantum
mechanical calculation \cite{wtyang,schaefer}
has become an indispensable research tool
and experimentalists have been increasingly relying on
computational results to interpret their experimental findings,
the practically used numerical methods by far
are often not accurate enough,
in particular, for complex systems.
This limitation is caused by the
inherent approximations adopted in the first-principles methods.
Because of computational cost, electron correlation has always
been a difficult obstacle for first-principles calculations.
Finite basis sets chosen in practical
computations are not able to cover entire physical space and this
inadequacy introduces further
inherent computational errors.
Effective core potential is frequently used to
approximate the relativistic effects, resulting inevitably in
errors for systems that contain heavy atoms.
The accuracy of a density-functional theory
(DFT) calculation is mainly determined by the
exchange-correlation (XC) functional
being employed~\cite{wtyang}, whose exact form is however unknown.
 Nevertheless, the results of first-principles
quantum mechanical calculation can
capture the essence of physics. For instance,
the calculated results, despite that their absolute values
may poorly agree with measurements, are usually of the same
tendency among different molecules as their experimental counterpart.
The quantitative discrepancy between
the calculated and experimental results
depends predominantly on the property of primary interest
and, to a less extent, also on other
related properties, of the material.
There exists thus a sort of quantitative relation
between the calculated and experimental results,
as the aforementioned approximations, to a large extent,
contribute to the systematic errors of specified first-principles methods.
Can we develop general ways to eliminate the systematic computational
errors and further to quantify the accuracies of numerical methods used?
It has been proven an extremely difficult task
to determine the calculation errors from the first-principles.
Alternatives must be sought.

We propose here a neural--network algorithm to determine the
quantitative relationship between the experimental data and the
first-principles calculation results. The determined relation will
subsequently be used to eliminate the systematic deviations of the
calculated results, and thus, reduce the numerical uncertainties.
Since its beginning in the late fifties, Neural Networks has been
applied to various engineering problems, such as robotics, pattern
recognition, speech, and etc.~\cite{PRNN,nature533} As the first
application of Neural Networks to quantum mechanical calculations
of molecules, we choose the standard heat of formation $\DelH$ and
standard Gibbs energy of formation $\DelG$ at 298.15 K as the
properties of interest.

\begin{figure*}
 \rotatebox{-90}{
\includegraphics[scale=0.7]{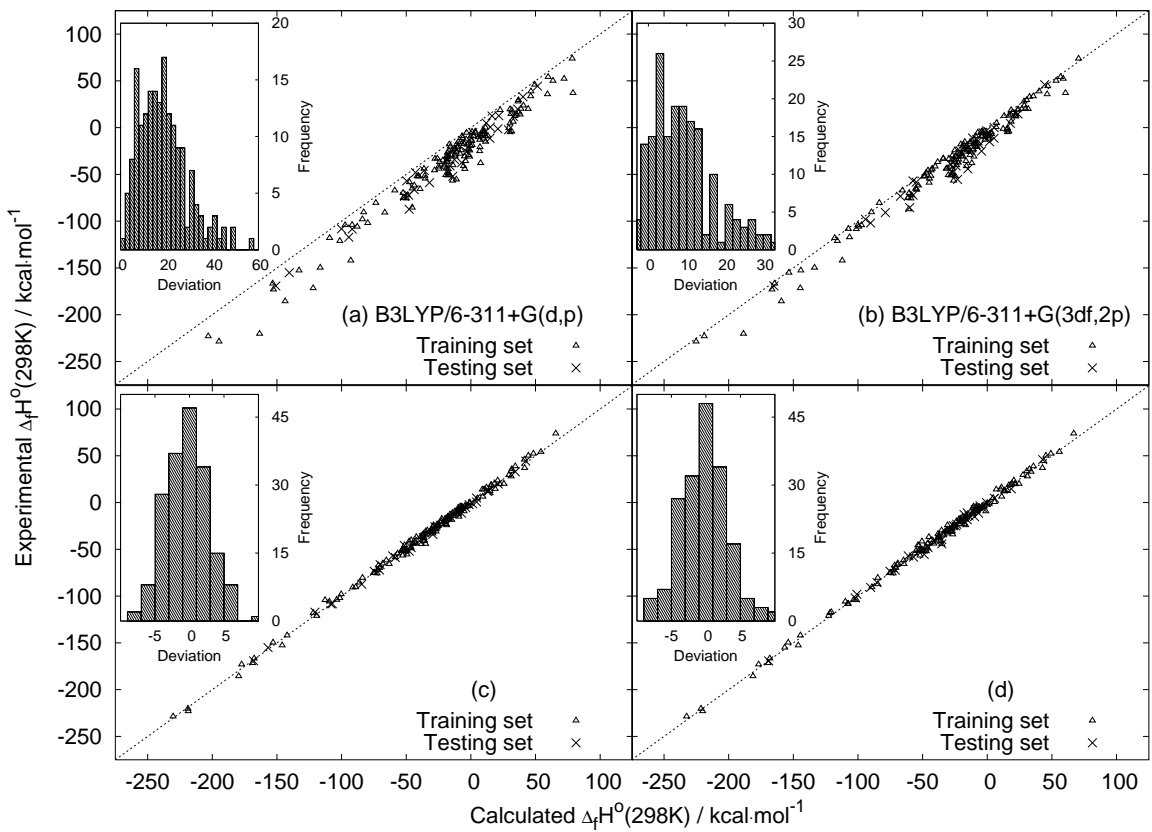}
    }
 \caption{\label{fig:wide1}Experimental $\DelH$ versus calculated $\DelH$
 for all 180 compounds. (a) and (b) are for raw B3LYP/6-311+G({\it d,p})
  and B3LYP/6-311+G(3{\it df},2{\it p}) results, respectively. (c) and (d)
  are for neural-network corrected B3LYP/6-311+G({\it d,p}) and
  B3LYP/6-311+G(3{\it df},2{\it p}) $\DelH$s, respectively.
  In (c) and (d), triangles are for the training set and crosses for the
  testing set. Inserts are the histograms for the differences between the
  experimental and calculated $\DelH$s. All values are in the units of
  kcal$\cdot$mol$^{-1}$.}
\end{figure*}

\begin{figure*}
 \rotatebox{-90}{
\includegraphics[scale=0.7]{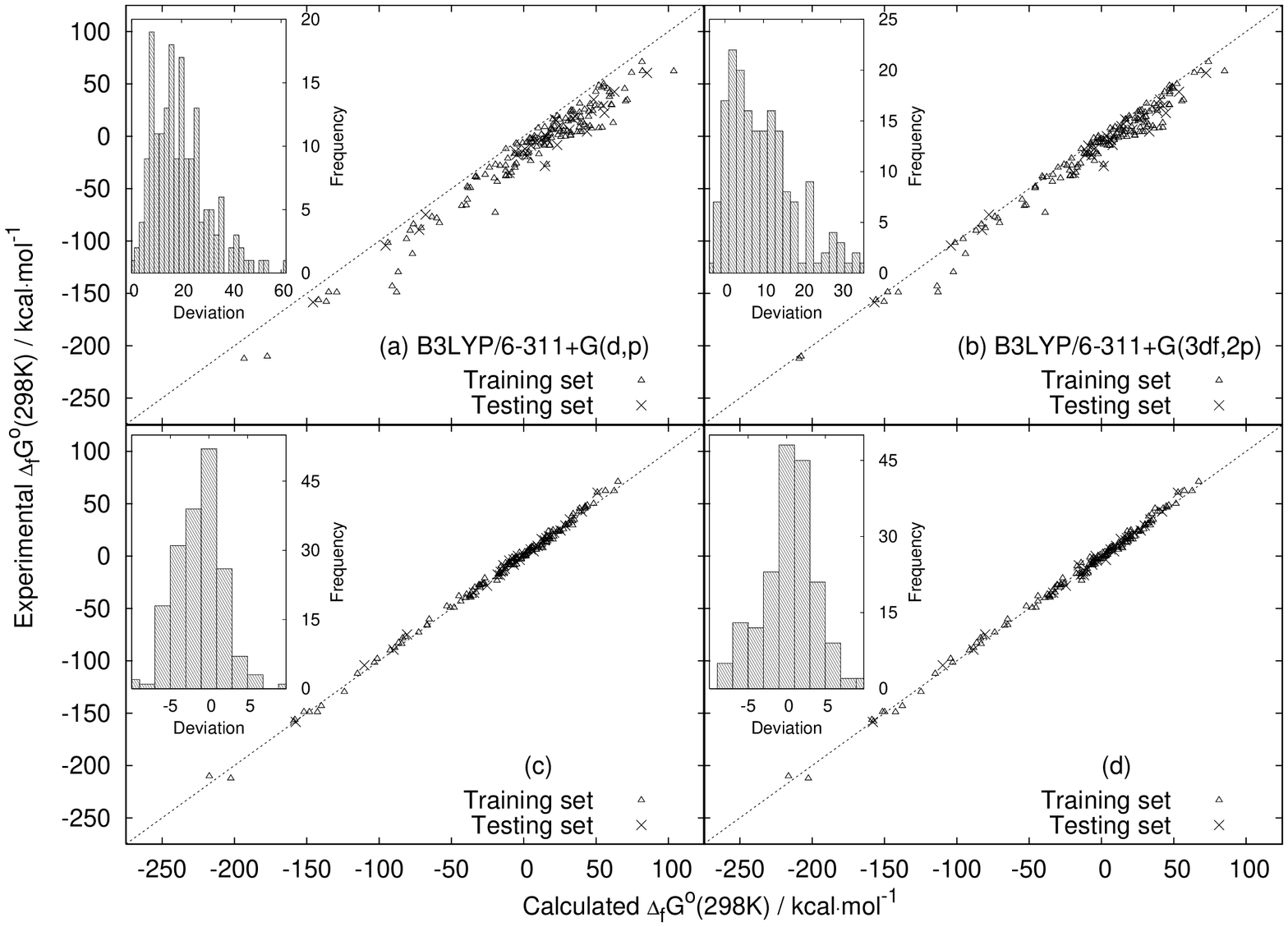}
    }
 \caption{\label{fig:wide2} Experimental $\DelG$ versus calculated $\DelG$
 for all 180 compounds. (a) and (b) are for raw B3LYP/6-311+G({\it d,p})
 and B3LYP/6-311+G(3{\it df},2{\it p}) results, respectively. (c) and (d)
 are for neural-network corrected B3LYP/6-311+G({\it d,p}) and
 B3LYP/6-311+G(3{\it df},2{\it p}) $\DelG$s, respectively. Legends, units
 and inserts are similar to those of Fig.~\ref{fig:wide1}.}
\end{figure*}

A total of 180 small- or medium-sized organic molecules, whose
$\DelH$ and $\DelG$ values are well documented in Refs.\
\onlinecite{McGraw,crchandbook,thermodata}, are selected to test
our proposed approach. The tabulated values of $\DelH$ and $\DelG$
in the three references differ less than 1.0
kcal$\cdotp$mol$^{-1}$ for same molecule. The uncertainties of all
$\DelH$ values are less than 1.0 kcal$\cdotp$mol$^{-1}$, while
those of $\DelG$s are not reported in Refs.\
\onlinecite{McGraw,crchandbook,thermodata}. These selected
molecules contain elements such as H, C, N, O, F, Si, S, Cl and
Br. The heaviest molecule contains 14 heavy atoms, and the largest
has 32 atoms. We divide these molecules randomly into the training
set (150 molecules) and the testing set (30 molecules). The
geometries of 180 molecules are optimized via B3LYP/6-311+G({\it
d,p})~\cite{g98} calculations and the zero point energies (ZPEs)
are calculated at the same level. The enthalpy and Gibbs energy of
each molecule are calculated at both B3LYP/6-311+G({\it d,p}) and
B3LYP/6-311+G(3{\it df},2{\it p}).~\cite{g98} B3LYP/6-311+G(3{\it
df},2{\it p}) employs a larger basis set than B3LYP/6-311+G({\it
d,p}). The unscaled B3LYP/6-311+G({\it d,p}) ZPE is employed in
the $\DelH$ and $\DelG$ calculations. The strategies in reference
\onlinecite{jcp97g2} are adopted to calculate $\DelH$ and $\DelG$.
The calculated $\DelH$ and $\DelG$ for B3LYP/6-311+G({\it d,p})
and B3LYP/6-311+G(3{\it df},2{\it p}) are compared to their
experimental counterparts in Figs.~\ref{fig:wide1} and
~\ref{fig:wide2}, respectively. The horizontal coordinates are the
raw calculated data, and the vertical coordinates are the
experimental values. The dashed lines are where the vertical and
horizontal coordinates are equal, {\it i.e.}, where the B3LYP
calculations and experiments would have the perfect match. The raw
calculation values are mostly below the dashed line, {\it i.e.},
most raw $\DelH$ and $\DelG$ are larger than the experimental
data. In another word, there are systematic deviations for both
B3LYP $\DelH$ and $\DelG$. Compared to the experimental
measurements, the root--mean--square (RMS) deviations for $\DelH$
($\DelG$) are 21.4 (22.3) and 12.0 (12.9) kcal$\cdotp$mol$^{-1}$
for B3LYP/6-311+G({\it d,p}) and B3LYP/6-311+G(3{\it df},2{\it p})
calculations, respectively. In Table~\ref{tab:table1} we compare
the B3LYP and experimental $\DelH$s for 10 of 180 molecules.
Overall, B3LYP/6-311+G(3{\it df},2{\it p}) calculations yield
better agreements with the experiments than B3LYP/6-311+G({\it
d,p}). In particular, for small molecules with few heavy elements
B3LYP/6-311+G(3{\it df},2{\it p}) calculations result in very
small deviations from the experiments. For instance, the $\DelH$
deviations for CH$_4$ and CS$_2$ are only -0.5 and 0.6
kcal$\cdot$mol$^{-1}$, respectively. Our B3LYP/6-311+G(3{\it
df},2{\it p}) calculation results are also in good agreements with
those of reference \onlinecite{jcp97g2} which employed a similar
calculation strategy except that their ZPEs were scaled by a
factor of 0.98 or 0.96 and their geometries were optimized at
B3LYP/6-31+G({\it d}). For large molecules, both
B3LYP/6-311+G({\it d,p}) and B3LYP/6-311+G(3{\it df},2{\it p})
calculations yield quite large deviations from their experimental
counterparts.
 \begin{table}
  \begin{center}
   \caption{\label{tab:table1}Experimental and calculated $\DelH$(298 K) for ten selected compounds (all data are in the units of kcal$\cdot$mol$^{-1}$)}
  \scriptsize{
  \begin{ruledtabular}
    \begin{tabular}{lrrrrrr}
         &  &\multicolumn{5}{c}{ Deviations (Theory-Expt.)}
 \\ \cline{3-7}
Molecules& Expt.\footnote{The experimental values were taken from
reference~\cite{thermodata}.} &
 DFT1\footnote{The deviations of calculated $\DelH$ by using B3LPY/6-311+G({\it d,p})
geometries, zero point energies and \\ $~~~~$enthalpies.} &
DFT1-NN\footnote{The deviations of calculated $\DelH$ by
B3LYP/6-311+G({\it{d,p}})-Neural Networks approach.} &
DFT2\footnote{The deviations of calculated $\DelH$ by using the
6-311+G({\it d,p}) geometries and zero point energies, and \\
$~~~~$the calculated enthalpies with 6-311+G(3{\it df},2{\it p})
basis.} & DFT2-NN\footnote{The deviations of calculated $\DelH$ by
B3LYP/6-311+G(3{\it df},2{\it p})-Neural Networks approach.} &
DFT3\footnote{The deviations were taken from \cite{jcp97g2}, where
the zero point energies were corrected by a scale factor.}
\\ \hline
CF$_2$O       & -152.9$\pm {0.4}$& 20.0 &6.9& 8.7  & 6.8 & 9.1  \\
CH$_2$Cl$_2$  &  -22.8$\pm{0.3}$ & 10.6 &3.6& 5.0  &  4.9& 4.6  \\
CH$_2$F$_2$   & -108.1$\pm {0.2}$& 8.0  &0.9& 0.6  & 0.6 & 0.0 \\
CH$_4$        &  -17.8$\pm {0.1}$& 1.1  &1.1& -0.5 & 1.0 &-1.6 \\
CS$_2$        &   27.9$\pm {0.2}$& 8.7  &3.3& 0.6  & 3.2 & 0.2 \\
C$_5$H$_{12}$ &  -35.1$\pm {0.2}$& 16.7 &-2.1&9.9  &-2.2 & -- \\
C$_5$H$_{12}$O&  -75.3$\pm {0.3}$& 23.2 &0.2& 14.0 &  0.1& -- \\
C$_6$H$_{14}$ & -41.1 $\pm {0.2}$& 25.1 &1.4& 17.0 & 1.5 & --  \\
C$_8$H$_{10}$ &  4.6  $\pm {0.3}$& 25.7 &0.5& 13.3 & 1.0 & --  \\
C$_9$H$_{12}$ & -2.3  $\pm {0.3}$& 31.3 &0.9& 17.6 & 1.8 & --  \\
  \end{tabular}
  \end{ruledtabular}
           }
  \end{center}
 \end{table}


\begin{figure}
\includegraphics[scale=0.5]{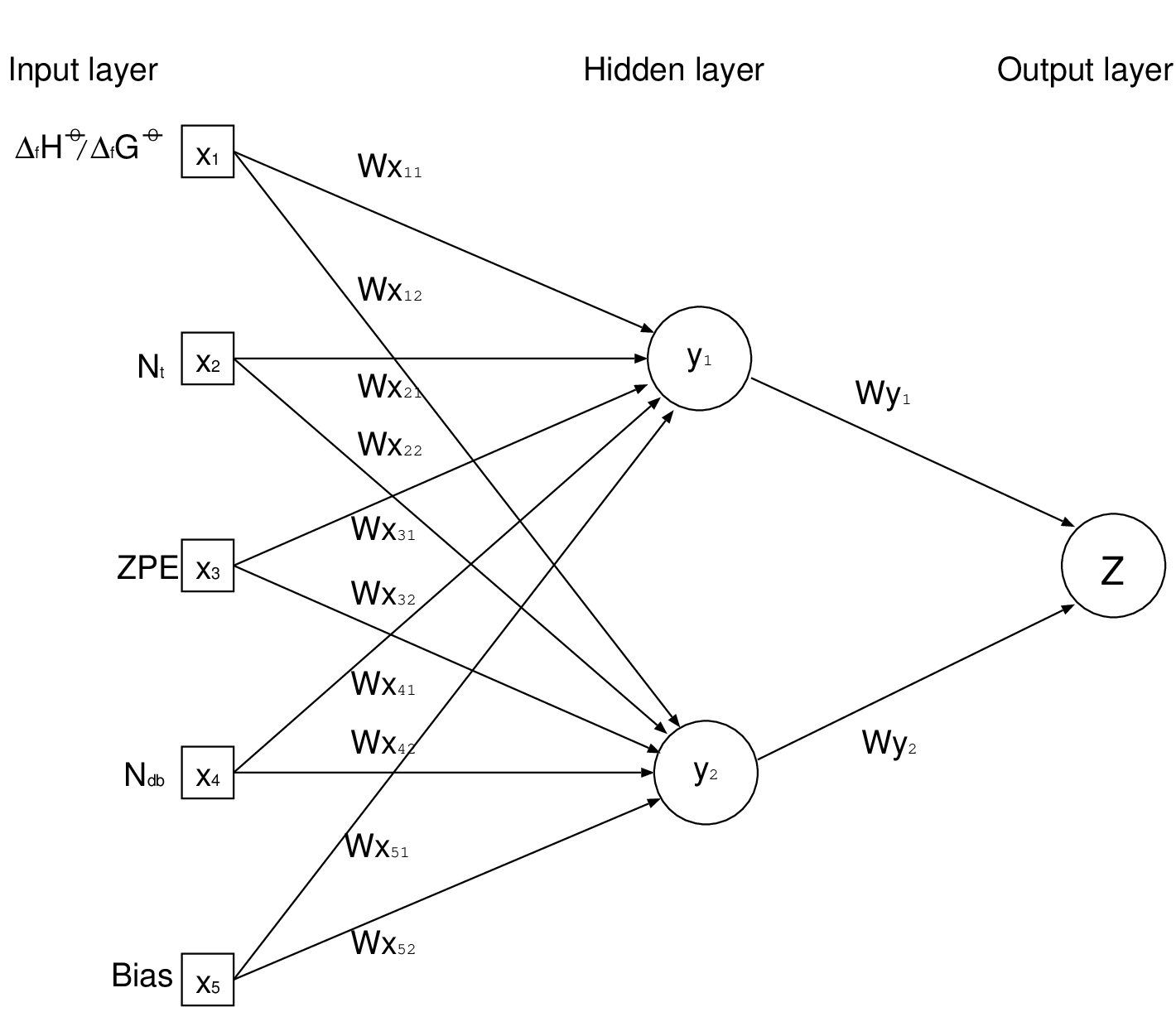}
 \caption{\label{fig:nnst} Structure of our neural network.}
\end{figure}

Our neural network adopts a three-layer architecture which has an
input layer consisted of input from the physical descriptors and a
bias, a hidden layer containing a number of hidden neurons, and an
output layer that outputs the corrected values for $\DelH$ or
$\DelG$ (see Fig.~\ref{fig:nnst}). The number of hidden neurons is
to be determined. The most important issue is to select the proper
physical descriptors of our molecules, which are to be used as the
input for our neural network. The calculated $\DelH$ and $\DelG$
contain the essence of exact $\DelH$ and $\DelG$, respectively,
and are thus obvious choices of the primary descriptor for
correcting $\DelH$ and $\DelG$, respectively. We observe that the
size of a molecule affects the accuracies of calculations. The
more atoms a molecule has, the worse the calculated $\DelH$ and
$\DelG$ are. This is consistent with the general observations in
the field.~\cite{jcp97g2} The total number of atoms $N_t$ in a
molecule is thus chosen as the second descriptor for the molecule.
ZPE is an important parameter in calculating $\DelH$ and $\DelG$.
Its calculated value is often scaled in evaluating $\DelH$ and
$\DelG$,~\cite{jcp97g2} and it is thus taken as the third physical
descriptor. Finally, the number of double bonds, $N_{db}$, is
selected as the fourth and last descriptor to reflect the chemical
structure of the molecule.

To ensure the quality of our neural network, a cross-validation
procedure is employed to determine our neural
network.~\cite{rbfnn} We divide further randomly 150 training
molecules into five subsets of equal size. Four of them are used
to train the neural network, and the fifth to validate its
predictions. This procedure is repeated 5 times in rotation. The
number of neurons in the hidden layer is varied from 2 to 10 to
decide the optimal structure of our neural network. We find that
the hidden layer containing two neurons yields best overall
results. Therefore, the 5-2-1 structure is adopted for our neural
network as depicted in Fig.~\ref{fig:nnst}. The input values at
the input layer, $x_1$, $x_2$, $x_3$, $x_4$ and $x_5$, are scaled
$\DelH$ (or $\DelG$), $N_t$, ZPE, $N_{db}$ and bias, respectively.
The bias $x_5$ is set to 1. The weights $\{ Wx_{ij}\}$s connect
the input layer $\{x_i\}$ and the hidden neurons $y_1$ and $y_2$,
and $\{ Wy_{j} \}$s connect the hidden neurons and the output Z
which is the scaled $\DelH$ or $\DelG$ upon neural-network
correction. The output Z is related to the input $\{x_i\}$ as
\begin{eqnarray}
Z = \sum_{j=1,2}Wy_{j}~ Sig(\sum_{i=1,5} Wx_{ij}~ x_i),
\end{eqnarray}
where $Sig(v) = {1\over 1+exp(-\alpha v)}$ and $\alpha$ is a
parameter that controls the switch steepness of Sigmoidal function
$Sig(v)$. An error back-propagation learning
procedure~\cite{nature533} is used to optimize the values of
$Wx_{ij}$ and $Wy_{j}$($i=1,2,3,4,5$ and $j=1,2$). In
Figs.~\ref{fig:wide1}c,~\ref{fig:wide1}d,~\ref{fig:wide2}c
and~\ref{fig:wide2}d, the triangles belong to the training set and
the crosses to the testing set. Compared to the raw calculated
results, the neural-network corrected values are much closer to
the experimental values for both training and testing sets. More
importantly, the systematic deviations for $\DelH$ and $\DelG$ in
Figs.~\ref{fig:wide1}a,~\ref{fig:wide1}b,~\ref{fig:wide2}a
and~\ref{fig:wide2}b are eliminated, and the resulting numerical
deviations are reduced substantially. This can be further
demonstrated by the error analysis performed for the raw and
neural-network corrected $\DelH$s and $\DelG$s of all 180
molecules. In the inserts of Figs.~\ref{fig:wide1}
and~\ref{fig:wide2}, we plot the histograms for the deviations
(from the experiments) of the raw B3LYP $\DelH$s and $\DelG$s and
their neural--network corrected values. Obviously, the raw
calculated $\DelH$s and $\DelG$s have large systematic deviations
while the neural--network corrected $\DelH$s and $\DelG$s have
virtually no systematic deviations. Moreover, the remaining
numerical deviations are much smaller. Upon the neural-network
corrections, the RMS deviations of $\DelH$s ($\DelG$s) are reduced
from 21.4 (22.3) kcal$\cdot$mol$^{-1}$ to 3.1 (3.3)
kcal$\cdot$mol$^{-1}$ and 12.0 (12.9) kcal$\cdot$mol$^{-1}$ to 3.3
(3.4) kcal$\cdot$mol$^{-1}$ for B3LYP/6-311+G({\it d,p}) and
B3LYP/6-311+G(3{\it df},2{\it p}), respectively. Note that the
error distributions after the neural--network correction are of
approximate Gaussian distributions (see Figs.~\ref{fig:wide2}c and
~\ref{fig:wide2}d).
 Although the raw
B3LYP/6-311+G({\it d,p}) results have much larger deviations than
those of B3LYP/6-311+G(3{\it df}, 2{\it p}), the neural--network
corrected values of both calculations have deviations of the same
magnitude. This implies that it is sufficient to employ the
smaller basis set 6-311+G({\it d,p}) in our combined DFT
calculation and neural--network correction (or DFT-NEURON)
approach. The neural--network algorithm can correct easily the
deficiency of a small basis set. Therefore, the DFT-NEURON
approach can potentially be applied to much larger systems. In
Table~\ref{tab:table1} we also list the neural--network corrected
$\DelH$s of the 10 molecules. The deviations of large molecules
are of the same magnitude as those of small molecules. Unlike
other quantum mechanical calculations that usually yield worse
results for larger molecules than for small ones, the DFT-NEURON
approach does not discriminate against the large molecules.

Analysis of our neural network reveals that the weights connecting
the input for $\DelH$ or $\DelG$ have the dominant contribution in
all cases. This confirms our fundamental assumption that the
calculated $\DelH$ ($\DelG$) captures the essential values of
exact $\DelH$ ($\DelG$). The input for the second physical
descriptor, $N_{t}$, has quite large weights in all cases. In
particular, when the smaller basis set 6-311+G({\it d,p}) is
adopted in the B3LYP calculations, $N_t$ has the second largest
weights. It is found that the raw $\DelH$ and $\DelG$ deviations
are roughly proportional to $N_{t}$, which confirms the importance
of $N_{t}$ as a significant descriptor of our neural network. The
bias contributes to the correction of systematic deviations in the
raw calculated data, and has thus significant weights. When the
larger basis set 6-311+G(3{\it df},2{\it p}) is used, the bias has
the second largest weights for all cases. ZPE has been often
scaled to account for the discrepancies of $\DelH$s or $\DelG$s
between calculations and experiments,~\cite{jcp97g2} and it is
thus expected to have large weights. This is indeed the case,
especially when the smaller basis set 6-311+G({\it d,p}) is
adopted in calculations. In all cases the number of double bonds,
$N_{db}$, has the smallest but non-negligible weights. In
Table~\ref{tab:table2} we list the values of $\{Wx_{ij}\}$ and
$\{Wy_{j}\}$ of the two neural networks for correcting $\DelG$s of
B3LYP/6-311+G({\it d,p}) and B3LYP/6-311+G(3{\it df},2{\it p})
calculations.
\begin{table}
   \caption{\label{tab:table2}Weights of DFT-Neural Networks for $\DelG$}
   \begin{ruledtabular}
    \begin{tabular}{lrrrrrrrrrrrrrrrrr}
& &&&\multicolumn{5}{c}{DFT1-NN\footnote{DFT1-NN refers
B3LYP/6-311+G({\it d,p})-Neural Networks approach.}}
&&&&&\multicolumn{5}{c}{DFT2-NN\footnote{DFT2-NN refers
B3LYP/6-311+G(3{\it df},2{\it p})-Neural Networks approach.}}\\
\cline{5-10} \cline{13-18} Weights  &&  & &y$_1$&&& &y$_2$
&&&&&y$_1$ &&&&y$_2$ \\ \hline
Wx$_ {1j}$        &&& &0.78 &&&&-0.72&&&&&0.83 &&&&-0.73 \\
Wx$_{2j}$         &&& &-0.60&&&&0.02 &&&&&-0.30&&&&0.02  \\
Wx$_{3j}$         &&& &0.44 &&&&0.02 &&&&&0.18 &&&&0.02  \\
Wx$_{4j}$         &&& & 0.07&&&& 0.24&&&&& 0.05&&&&0.17  \\
Wx$_{5j}$         &&& &-0.42&&&&-0.04&&&&&-0.46&&&&0.01  \\
Wy$_{j}$          &&& &1.48 &&&&-0.57&&&&&1.44 &&&&-0.47 \\
  \end{tabular}
\end{ruledtabular}
 \end{table}

Our DFT-NEURON approach has a RMS deviation of $\sim$3
kcal$\cdot$mol$^{-1}$ for the 180 small- to medium-sized organic
molecules. This is slightly larger than their experimental
uncertainties.~\cite{McGraw,crchandbook,thermodata} The physical
descriptors adopted in our neural network, the raw calculated
$\DelH$ or $\DelG$, the number of atoms $N_t$, the number of
double bonds $N_{db}$ and the ZPE are quite general, and are not
limited to special properties of organic molecules. The DFT-NEURON
approach developed here is expected to yield a RMS deviation of
$\sim$3 kcal$\cdot$mol$^{-1}$ for $\DelH$s and $\DelG$s of any
small- to medium-sized organic molecules. G2 method~\cite{jcp97g2}
results are more accurate for small molecules. However, our
approach is much more efficient and can be applied to much larger
systems. To improve the accuracy of the DFT-NEURON approach, we
need more and better experimental data, and possibly, more and
better physical descriptors for the molecules. Besides $\DelH$ and
$\DelG$, the DFT-NEURON approach can be generalized to calculate
other properties such as ionization energy, dissociation energy,
absorption frequency, band gap and etc. The raw first-principles
calculation property of interest contains its essential value, and
is thus always the primary descriptor. Since the raw calculation
error accumulates as the molecular size increases, the number of
atoms $N_t$ should thus be selected as a descriptor for any
DFT-NEURON calculations. Additional physical descriptors should be
chosen according to their relations to the property of interest
and to the physical and chemical structures of the compounds.
Others have used Neural Networks to determine the quantitative
relationship between the experimental thermodynamic properties and
the structure parameters of the molecules.~\cite{rbfnn} We
distinct our work from others by utilizing specifically the
first-principles methods and with the objective to improve quantum
mechanical results. Since the first-principles calculations
capture readily the essences of the properties of interest, our
approach is more reliable and covers much a wider range of
molecules or compounds.

To summarize, we have developed a promising new approach to improve
the results of first-principles quantum mechanical
calculations and to calibrate their uncertainties.
The accuracy of DFT-NEURON approach can be systematically
improved as more and better experimental data are available.
As the systematic deviations caused by small basis sets and less sophisticated
methods adopted in the calculations can be easily corrected by Neural Networks,
the requirements on first-principles calculations are modest. Our approach
is thus highly efficient compared to much more sophisticated first-principles methods of
similar accuracy, and more importantly, is expected to be applied
to much larger systems. The combined first-principles calculation and
neural-network correction approach developed in this work is potentially
a powerful tool in computational physics and chemistry, and
may open the possibility for first-principles methods to be
employed practically as predictive tools in materials research and design.

We thank Prof. YiJing Yan for extensive discussion on the subject and generous
help in manuscript preparation.
Support from the Hong Kong Research Grant Council (RGC) and the
Committee for Research and Conference Grants (CRCG) of the
University of Hong Kong is gratefully acknowledged.

\end{document}